# Temperature dependent vibrational spectra in non-crystalline materials: application to hydrogenated amorphous silicon

I.M. Kupchak, F. Gaspari[*], A.I. Shkrebtii, J. Perz

Faculty of Science, University of Ontario Institute of Technology,

2000 Simcoe Street North, Oshawa, ON, L1H 7K4 Canada

**Abstract**

We present a novel approach for parameter-free modeling of the structural, dynamical and electronic properties of non-crystalline materials based on *ab-initio* Molecular Dynamics, improved signal processing technique and computer visualization. The method have been extensively tested by investigating hydrogen and silicon dynamics in hydrogenated amorphous silicon (a-Si:H). By comparing the theoretical and experimental vibrational spectra we demonstrate how to relate vibrational properties to the structural stability, bonding and hydrogen diffusion. We extracted microscopic characteristics that cannot be obtained by other techniques, namely hydrogen migration and related bond switching, dangling bond passivation, low hydrogen activation energy, and a-Si:H stability in general, and we show, via the analysis of a test case, that our method provides a rigorous and realistic description of non-crystalline materials. We also demonstrate that this method offers the possibility of accessing other important macroscopic characteristics of amorphous silicon and can be used to model all the aspects of a-Si:H dynamics, including the detrimental Staebler-Wronski effect.

---

[*] *Corresponding author:* franco.gaspari@uoit.ca

**Introduction**

Molecular dynamics (MD) has become one of the more powerful and more frequently used tools for the correlation of the microscopic characteristic of materials with their macroscopic properties, observed experimentally. Non-crystalline materials are important systems that can be investigated by MD. Some relevant macroscopic properties of non-crystalline systems are, for instance, radial distribution function, diffusion, vibrational spectra, optical gap, *etc*. Vibrational techniques are one of the main sources of experimental information that helps to understand the properties of non-crystalline materials [1-3]. In spite of the importance of theoretical calculations of the vibrations, the critical issue remaining is how to properly theoretically calculate the vibrational spectra. Baroni and Giannozzi [4] first developed a numerical approach to model the vibrational spectra from the first principles a decade ago, but this is only efficient for crystalline systems.

We have developed and applied the method to Hydrogenated Amorphous Silicon (a-Si:H), a material which has been the subject of intensive investigation for at least 30 years, and for which there exists an extensive literature covering all of its most important properties.

In this paper we focus primarily on establishing correlation of the hydrogen vibrations at room temperature and below with a-Si:H properties by comparing theory and experiment. This comprehensive first-principles molecular dynamics (MD) simulation of the microscopic processes in a-Si:H, focuses mainly on the vibrational frequencies as a signature of the macroscopic material properties. In particular, we correlate for the first time the vibrational spectra of a-Si:H, hydrogen migration and radial distribution function with the preparation conditions, annealing and hydrogen concentration. We will also show that our novel approach enables a more realistic description, a deeper understanding of these complex materials, and a more accurate evaluation of the macroscopic properties.

Although the properties of amorphous Si have been intensively studied for a few decades, a number of fundamental issues remain unresolved. Microscopic atom dynamics, for instance, influences atomic structure, chemical bonding, diffusion and vibrations, and are difficult to study both experimentally and theoretically. However, the microscopic details of disordering, hydrogen migration and bonding within the amorphous silicon network is crucial for the understanding of a-Si:H, including the detrimental Staebler-Wronski (SW) effect [5], and for the improvement of the overall quality of the material. For instance, it was not clear how to experimentally predict the stability of a-Si-H films, grown at particular temperature and hydrogen concentration, with respect to SWE. Vibrational spectroscopies [6-8] demonstrated essential modification of the hydrogen behavior for different concentration and growth conditions. The vibrational spectra clearly indicate on the different processes inside a-Si:H, but these processes were not interpreted theoretically in terms of, *e.g.,* bonding, hydrogen dynamics and migration. The only way to access this on a microscopic level is to model the system numerically.

Many theoretical techniques have been employed to establish a realistic microscopic model of a-Si:H, including classical [9,10] and density-functional-based tight-binding Molecular Dynamics [11]. However, verification of the "realism" of the model has mostly been limited to the derivation of the radial distribution function (RDF) and its comparison with experimental data. Some authors [12,13] have simulated the dynamics of hydrogen diffusion, in particular as a way of analyzing possible mechanism behind the Staebler-Wronski effect. Others have calculated the density of states (DOS) and the vibrational properties using the tight-binding approximation [11,14]. Monte-Carlo method has also been attempted [15]. However, the classical MD is not accurate enough to describe the covalent bonding and forces in semiconductors. Monte-Carlo method, on the other hand, does not produce real atomic trajectories, required to study dynamical effects. Finally, tight-binding approach (even DFT-LDA based [11]) is not sufficiently transferable to the disordered systems. For the hydrogenated amorphous Si, an absence of translational symmetry, presence of non-saturated Si bonds, hydrogen switching between host Si atoms, significant bond inharmonicity and diffusion, the first principles or *ab-initio* molecular dynamics (AIMD) remains the most accurate approach, perfect for description of the interatomic potential, explicitly accounting for the inharmonicity. However, there are only a few AIMD simulations of a-Si:H [12,13,16,17], Unfortunately, the above results were mostly limited to only one H concentration and high temperatures. In addition, they did not address vibrational spectra.

The main role of hydrogen in amorphous Si is the passivation of the DBs to restore a proper energy gap and the semiconducting properties, thus enabling extensive application of a-Si:H in the microelectronics and the photovoltaic industry. Due to the importance of hydrogen, many experimental methods have been used to characterize the DBs passivation, bonding chemistry and related mechanisms of degradation of the material. Among the numerous experimental techniques used to study a-Si:H and the role of hydrogen, the Fourier Transform Infrared Spectroscopy (FTIR) is used extensively to analyze vibrational spectra of a-Si:H. Although FTIR represents one of the most common and powerful techniques (see ref. [18], sections 2.1 and 2.3), no microscopic links between the observed vibrational features of the hydrogen and the microscopic properties of a-Si:H were established.

We have investigated hydrogen vibrations, diffusion, and bonding in a-Si:H from the time dependent atomic trajectories. We have identified for the first time signatures of hydrogen instability in a-Si:H, its re-bonding, and formation of various complexes from the vibrational spectra, thus extracting theoretically various macroscopic properties of a-Si:H  from microscopic AIMD simulations. Finally, a comparison with the experimental data provided us wealth of information regarding the hydrogen diffusion, and the quality of the material.

**Theoretical formalism**

We used the Car-Parrinello *ab initio* molecular dynamics [19] to simulate a-Si:H with different hydrogen concentration (up to 20%). The Density Functional Theory (DFT) based MD used was implemented in the software package Quantum-Espresso [20]. To represent the interaction between the valence electrons and the ionic cores, we used the first-principles norm-conserving pseudopotential in Car-van Barth form [21]. The 64 Si atom supercell with hydrogen atoms has been used throughout the MD runs. Kohn-Sham orbitals were expanded in a plain wave basis set using an energy cut-off for the wave function expansion of 12 Ry, and Brillouin zone of the supercell lattice was sampled by the Γ-point. A few runs with higher cutoff up to 20 Ry have been performed to confirm that the simulations at 12 Ry is accurate enough to describe the dynamics of both H and Si atoms. Nosé thermostat [22] was used to control the temperature of the supercells. The vibrational frequencies have been extracted from the time dependent atomic trajectories obtained from an additional MD runs by switching Nosé thermostat off. We have found that the vibrational spectra of amorphous 64 Si supercell with H atoms contains too many frequencies (compared, *e.g.*, to the crystalline Si), including essential numerical noise. Apart from this, Hydrogen and Silicon diffusion as well as H-bond switching modify the vibrational frequencies during the long MD run, thus further complicating vibrational analysis. To overcome this difficulty we have developed and intensively tested a new approach that combines the signal analysis method MUSIC [23] and Fourier transform. MUSIC was used first to distinguish between the signal and noise contributions dividing MD run into a few intervals, in which the frequencies do not change essentially. Secondly, MUSIC was applied to recreating "noise free" trajectories of H atoms. Finally, using this as an input, the Fourier transform has been applied to calculate the vibrational spectra. This allowed to clearly separating all vibrational modes of the Si-H complexes observed.

As it is generally accepted DFT underestimates vibrations frequencies of Hydrogen atoms by about 10% [14]. To overcome this problem we have determined a frequency scaling factor for a-Si:H, performing MD runs for the silane gas ($SiH_4$) atoms as well. Comparing the calculated and experimental frequencies for silane, we have found the factor of 1.13 has to be used at 12 Ry to correct the theoretical vibrational spectra.

We started MD with a crystalline supercell containing 64 silicon atoms. 4, 6, 8, 10 and 12 hydrogen atoms were randomly added into the crystal and geometrically optimized to zero temperature with the force tolerance threshold of $10^{-4}$ atomic units. The Si positions in the new structures were slightly distorted from the initial crystal geometry.

Next, we prepared a number of amorphous Si:H systems with different Hydrogen content by melting and subsequent slow cooling. The equations of motion were integrated using Verlet algorithm with a discrete time step of 0.24 fs. The samples were melted at 3000K (some MD runs were performed at 5000K) for 20000 MD steps (~5 ps), with temperature control via Nosé thermostat. After that we typically equilibrated the liquid system for 5-10 ps. The radial distribution functions indicate that we have well-equilibrated distribution of the atoms in the liquids. Nosé thermostat assisted cooling with different rate followed the melting. We were able to achieve a slowest cooling rate up to $4\times10^{13}$ K/sec (5 K per each 300 steps) for all samples, the longest annealing time (about 50 ps with annealing) and equilibration compared to the previous AIMD simulations (see, for instance, [12,17]). To recreate different experimental situations, we used a higher quenching rate of $7\cdot10^{13}$ K/s (5 K per each 500 MD steps) for $Si_{64}$ and $Si_{64}H_8$ samples as well.

After the cooling, the thermostat was switched off and 40000 MD steps (10 ps) were used to equilibrate the systems again and collect statistical data. In addition, we have minimized the total energy of a few amorphous structures with different hydrogen content to investigate the equilibrium lattice constant for the a-Si:H. We have found that the optimized bulk constant is close to that of crystalline Si. On the other hand, the equilibrium constant slightly depends on the particular amorphous network geometry in a random way. Therefore we have decided to use the density of the crystalline Si for the a-Si:H supercells in our simulations.

As a result of the extensive and long MD simulations for different H concentrations we were able to recreate all the main structural Si-H units, which have been suggested to from experimental measurements. These include Si-H, Si-$H_2$, Si-$H_3$ complexes and $H_2$ molecules inside amorphous Si network and their time evolution. Finally recrystallization of the amorphous Si has been achieved numerically for the first time. Dynamical trajectories before and after annealing include explicitly bond inharmonicity, amorphous network modifications due to Si and H diffusions, and bond switching or H bond sharing with a few Si atoms. For instance, the MD temperature change from 100K to 300K produces red shift of the Si-H stretching frequencies by a factor of 20-50 $cm^{-1}$. AIMD allowed realistic visualization of the above processes as well. The atomic trajectories of the various a-Si:H structures have been used to produce an animated movie, where all the atomic dynamics within the a-Si:H structure are shown. Finally the calculated vibrational spectra were correlated to the radial distribution function, separated into contributions from different atoms and Si-H complexes.

**Results**

We report the AIMD results for the vibrational frequencies of Si—H bonds for different hydrogen content. The infrared vibrational spectra are correlated to different Si-H bonding configuration [6], with the two main features centered at about 600 $cm^{-1}$ (bending modes) and between 2000 and 2100 $cm^{-1}$ (stretching modes). Other characteristic features are the scissors and bending peaks (also referred to as doublet peak) between 800 and 900 $cm^{-1}$ (generated by Si—$H_2$ and Si—$H_3$ bonds). Furthermore, the 2000 $cm^{-1}$ feature includes a variety of Si—H bonds, which can be distinguished by a proper deconvolution of the peak. For instance, Si—H monohydrides give rise to a signal at 2000 $cm^{-1}$, while monohydrides clusters, dihydride and polyhydrides show an absorption peak closer to 2100 $cm^{-1}$.

Fig. 1 shows a frame from an animation created from a $Si_{64}H_{10}$ structure (corresponding to 13.5 at% Hydrogen concentration). The radial distribution function for this structure is shown in Fig. 2.

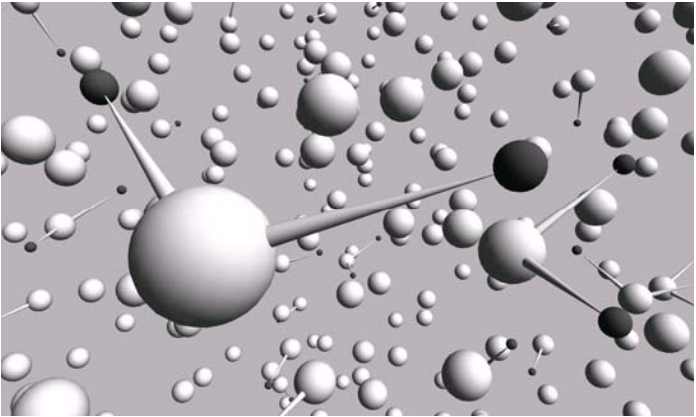

Fig. 1. a-Si:H structure obtained from MD simulation for a unit cell of 64 Si atoms and 10 H atoms. Si atoms are represented by the bigger, light colored spheres. H atoms are shown by smaller, black spheres. Only the bonds between Si atoms and H atoms are shown. Note the 2 dihydride bonds at the forefront of the picture.

As shown in Fig. 1, we were able to reproduce two important bonding configuration in this particular structure, i.e., six monohydride (Si—H) bonds for six of the Hydrogen atoms, and two dihydride (Si—$H_2$) bonds for the other four. The RDF shown in Fig. 2 indicates that this structure is indeed amorphous, and reproduces closely the RDF measured experimentally [24].

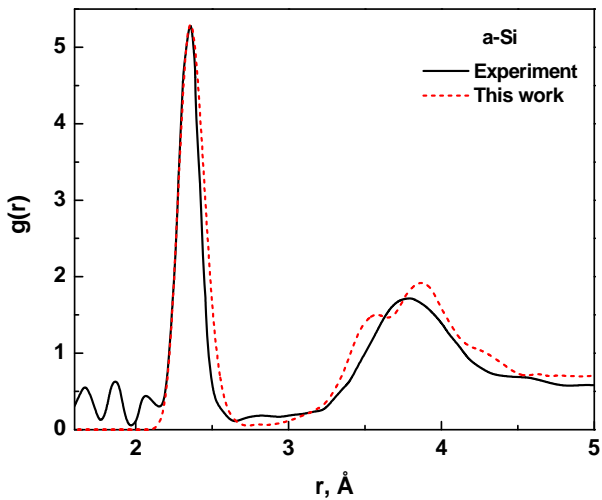

Fig. 2. The experimental RDF g(r) of annealed a-Si (solid, black line), and calculated (red, dashed line). Experimental data are taken from Laaziri *et al.*, *Phys. Rev. Lett.* **82**, 3460 (1999). Note: the experiment was carried out at 10K, while MD was at 300K, resulting in the widening of the peak.

In Fig. 3 we show the stretching vibrational modes calculated using our method for the structure with all the 10 Hydrogen atoms (black, solid line), for the same structure where the contribution to the vibrational frequencies from the 4 Hydrogen atoms bonded in a dihydride structure have been removed (red, long-dashed line), and again for the same structure where only the dihydrides frequencies are present (blue, short-dashed line). Other Hydrogen associated modes (i.e. wagging, bending, etc.) are shown in Fig. 4.

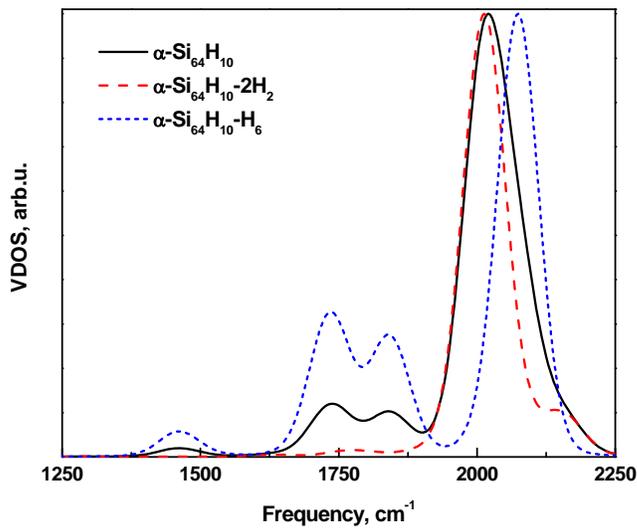

Fig. 3. Hydrogen vibrations, stretch mode only for a-$Si_{64}H_{10}$ system. The intensities have been normalized. We show all Hydrogen associated stretching vibrational modes (black, solid), dihydride modes (blue, short dash) and monohydride modes (red, long dash)

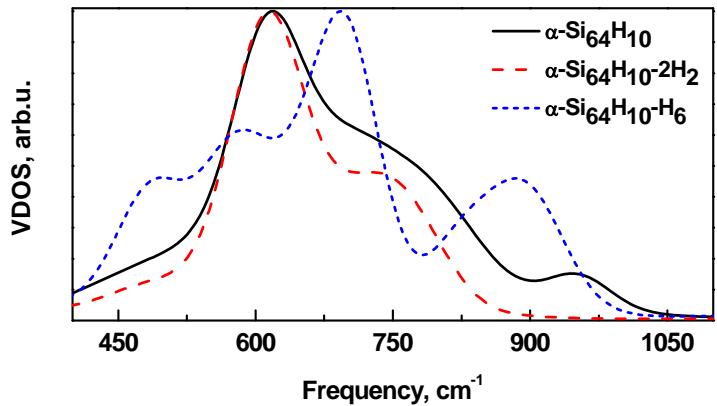

Fig. 4. Hydrogen vibrations, bending, wagging and rocking modes for a-$Si_{64}H_{10}$ system. Color codes are the same as for Fig. 3.

As can be seen, we were successful in reproducing the "standard" experimental observation associated with FTIR, including the Hydrogen stretching modes between 2000 cm$^{-1}$ and 2100 cm$^{-1}$. When the Hydrogen is bonded exclusively as a dihydride (blue, short-dashed line) the stretching mode is observed at about 2100 cm$^{-1}$, while monohydride bonds give rise to a peak at 2000 cm$^{-1}$, consistent with experimental observations. The effect of the dihydride modes is evident also for the lower frequency region, between 500 cm$^{-1}$ and 1000 cm$^{-1}$. Following Lucovsky *et al.* [6], the following infrared active vibrational modes can be associated with dihydride bonds: symmetric and asymmetric stretch (at ~ 2100 cm$^{-1}$), bend-scissors (~ 900 cm$^{-1}$), wag (~ 850 cm$^{-1}$) and rock (~ 620 cm$^{-1}$). All these modes are present for the structure with only dihydrides and for the overall structure, while they disappear for the monohydride structure, where they are replaced by the 2000 cm$^{-1}$ peak in the higher frequency region. A smaller feature can be seen in Fig. 3 between 1500 and 1800 cm$^{-1}$. We discuss the origin of this feature below.

Figs. 5a and 5b represent two frames, about 2.18 fs apart, of an AIMD generated movie showing the dynamics of the oscillations of the atoms in a $Si_{64}H_{12}$ structure at 300K. The hydrogen atom, small black sphere in the forefront, oscillates between two neighboring Si atoms. The H-atom remains bonded to a Si-site for an average of 15 fs. We were able to isolate the contributions to the frequencies from the "jumping" H-atom, to allow probing the frequencies of this type of bonding. Fig. 6 shows the vibrational frequencies for the above structure.

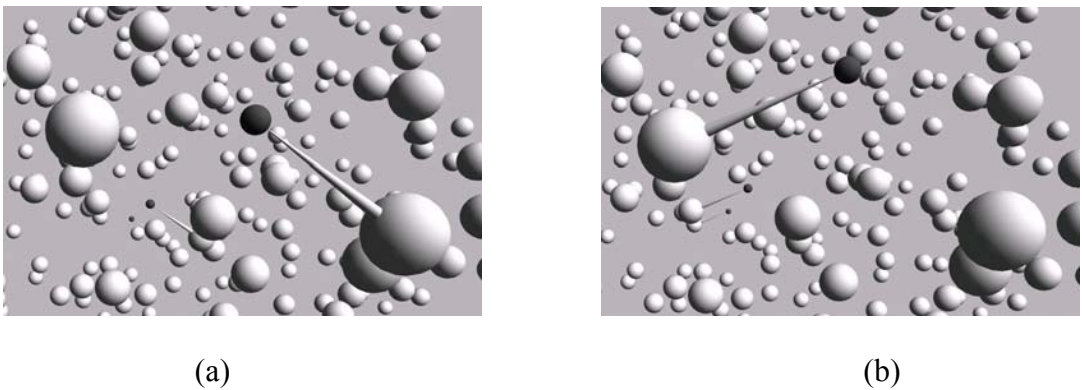

(a)                          (b)

Fig. 5. Successive frame of an AIMD generated movie showing the "jumps" of an H-atom between two neighboring Si atoms. The three atoms are in the forefront of the pictures.

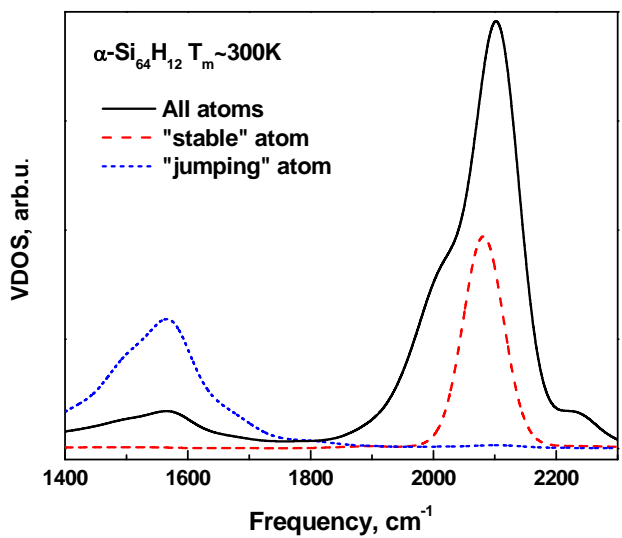

Fig. 6. Hydrogen stretching vibrations, for a-$Si_{64}H_{12}$ system. The blue line represents the vibrations of the "jumping" H-atom.

Once again, we observe a signature between 1500 and 1800 cm$^{-1}$, which appears to be related to the presence of a metastable Si—H bond. Darwich *et al.* [7], using infrared transmission spectroscopy (IRT) and infrared ellipsometry, observed a new, metastable feature at ~1730 cm$^{-1}$ during light-soaking, which they attributed to a three centre Si—H—Si bond (TCB). The authors remark that the theoretical calculation for the IRT frequency of such a complex had been a controversial issue, with the expected values lying between 800 and 1950 cm$^{-1}$, and proposed that in fact the band at 1730 cm$^{-1}$ represents the stretching mode of the TCB. In our simulations, we in fact observe a feature associated with a TCB type of bond between 1500 and 1800 cm$^{-1}$.

Our AIMD results confirm Darwich's claim, within experimental error. The decrease in the vibrational frequency with respect to that of a stable mono-hydride bond is due to the sharing of the Hydrogen electron density between two Si atoms. This decreases the Si-H bond strength, increases the bond length and reduces the vibrational amplitude   Therefore, the band in the 1500-1800 cm$^{-1}$ region is the signature of all Hydrogen metastable bonds, including the TCB bond, with variations in the frequency due to the different overlap between the H and the Si electron wave functions.

**Conclusions**

We have presented a novel approach for the parameter-free modeling of the structural, dynamical and electronic properties of non-crystalline materials based on *ab-initio* Molecular Dynamics.

We have shown that this approach enables a more realistic description of the macroscopic properties of non-crystalline materials. We have tested this method on hydrogenated amorphous silicon, and have found that our derivation of the macroscopic vibrational spectra agree very well with the experimental data. Furthermore, we have also investigated the test case of a three-centered Si—H—Si bond, first observed experimentally by Darwich *et al.* We show that our prediction of the vibrational modes of this configuration agrees with the experimentally observed value, and improve on previous AIMD analysis presented by Su *et al.* for the same bond structure. This novel approach can be applied to extract other fundamental macroscopic properties from the microscopic analysis, including the dynamic processes responsible for the Staebler-Wronski effect in a-Si:H.

The research was supported by the Centre for Materials and Manufacturing/Ontario Centres of Excellence (OCE/CMM) for the *Sonus/PV Photovoltaic Highway Traffic Noise Barrier* project and Shared Hierarchical Academic Research Computing Network (SHARCNET)